\documentstyle[twoside, epsf]{article}

\input ibvs2.sty

\begin{document}
\IBVShead{5813}{00 Month 200x}

\IBVStitle{H$\alpha$ observations of $\zeta$ Tauri}

\IBVSauth{E. Pollmann$^1$, Th. Rivinius$^2$}

\IBVSinst{Emil-Nolde-Str.\ 12, 51375 Germany}
\IBVSinst{ESO-Chile, Casilla 19001, Santiago de Chile}

\SIMBADobjAlias{zeta Tau}{HD 37202}
\IBVSkey{spectroscopy}
\IBVSabs{We report H$\alpha$ observations of $\zeta$\,Tauri, taken between
  late 2000 and early 2006.}

\begintext

\section{Introduction}
Be stars are well known to be variable on virtually all timescales, reaching
from minutes to dozens of years. For the study of the latter, long term data
collections as homogeneous as possible are necessary.

The professional astronomer, however, is often hampered in the study of
intermediate- to long-term time scale processes as in Be stars. The reasons
are the observational practices usually employed at professional
observatories, which typically are not suited for observing a bright object
with execution times of a few minutes only about every other week for several
seasons; as well as the funding timescales, making it hard to start the
collection of a long-term database that does not promise a significant number
of publications within the first few years.

On the other hand, the interpretation of time-limited observations with
professional resources, such as interferometers, polarimeter, or
high-resolution spectrographs, in almost all cases can profit from the
knowledge of the disc state in the course of the long-term evolution.

The problems in long-term data acquisition for the professional astronomer,
however, open a promising field for the dedicated amateur.  Amateur
spectrographs at relatively small telescopes of about 20\,cm diameter,
equipped with CCD-detectors meanwhile reach resolution powers well above
10\,000 and and are sensitive enough to reach many of the brighter Be stars.
This work describes a database worth of more than five years of observation of
the Be star $\zeta$ Tau.

The observational data will be made available online together with this
communication.

\IBVSfig{21cm}{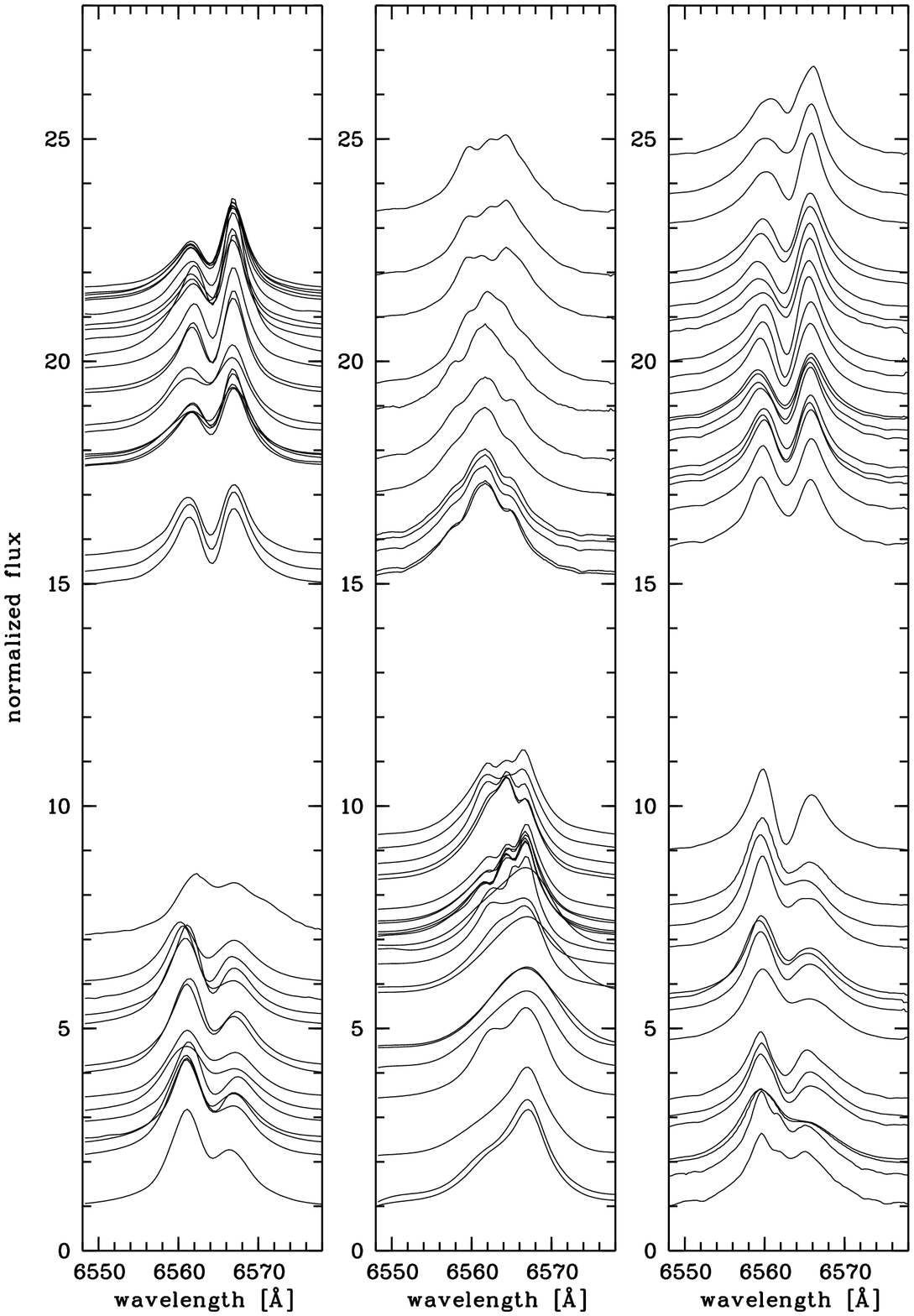}{All H$\alpha$ profiles measure from late
  2000 to early 2006. The vertical offset of the profiles is
  proportional to time and corresponds to 25 days per continuum
  unit. The lowermost spectra date from Nov.\ 1, 2000 (left), Sep.\ 9,
  2002 (middle) and Aug.\ 23, 2004 (right), respectively.}

\section{Observations}

$\zeta$ Tau is a well known frequently observed object.  Observations of the
H$\alpha$ emission line reach back many decades. This work amends those
series by the results of H$\alpha$-observations taken between late 2000 and
early 2006, i.e.\ six full observing seasons. All observations were made with
a 20\,cm Schmidt-Cassegrain telescope.  From Nov.\ 2000 to Apr.  2003, a
slitless prism-spectrograph with a dispersion of 43\,\AA/mm was used
($R\approx8000$), from Sep. 2003 to Apr.\ 2006 a slitless grating one with a
dispersion of 27\,\AA/mm and $R\approx14\,000$.

The spectra were normalized by hand-selecting a number of continuum points
through out the spectrum from 6500 to 6700\,\AA\ and then applying a spline
fit through those points.  The wavelength calibration was derived using
telluric features in the region of H$\alpha$, reaching an accuracy of about
0.1\AA\ on those features when compared to wavelengths derived with high
resolution instruments (telluric wavelengths measured with UVES were kindly
provided by R.\ Hanuschik, priv.\ comm.).

The H$\alpha$ spectra obtained by EP will be published electronically together
with this communication in the form of ASCII tables. The first column of each
table is holding the wavelength, while the first row notes the Julian date
(minus 2\,400\,000) at mid exposure.

\subsection{Equivalent width}
In the normalized and calibrated spectra, the H$\alpha$ equivalent width was
measured by integrating the normalized spectrum in the range from 6520 to
6600\,\AA. Comparison of the data presented here with quasi-simultaneous
spectra taken by Rivinius et al. (2006) confirm the scientific reliability of
the present data, both in terms of profile shape (see Fig.~1 vs.\ Rivinius et
al.) and equivalent width (see Figs.~2 and 3).

In theory, the measured equivalent width should be independent from
dispersion. In practice, this is typically not the case, however: spectra
with lower resolution, i.e. the ones with 43\AA/mm, differ systematically
from higher resolution data. In our observations, this can be seen from the
available quasi-simultaneous observations with professional instruments. We
attempt no correction of this effect, but rather point out its existence in
order not to over-interpret the data.

In general, the accuracy of amateur instruments for measuring equivalent
widths currently is hardly better than about 5\,\%. 

To check the accuracy obtained, both for the equivalent width and the peak
height ratio of the emission, a series of observations of standard stars was
obtained in three nights, 8h worth of observations in total. For both
quantities, the RMS-error of the individual measurements in a single night was
below 3\,\%. No correction for the contamination due to telluric vapour lines
to the total EW was attempted, as the effect is, with about 1\,\%, well below
the measuring accuracy.

\IBVSfig{9cm}{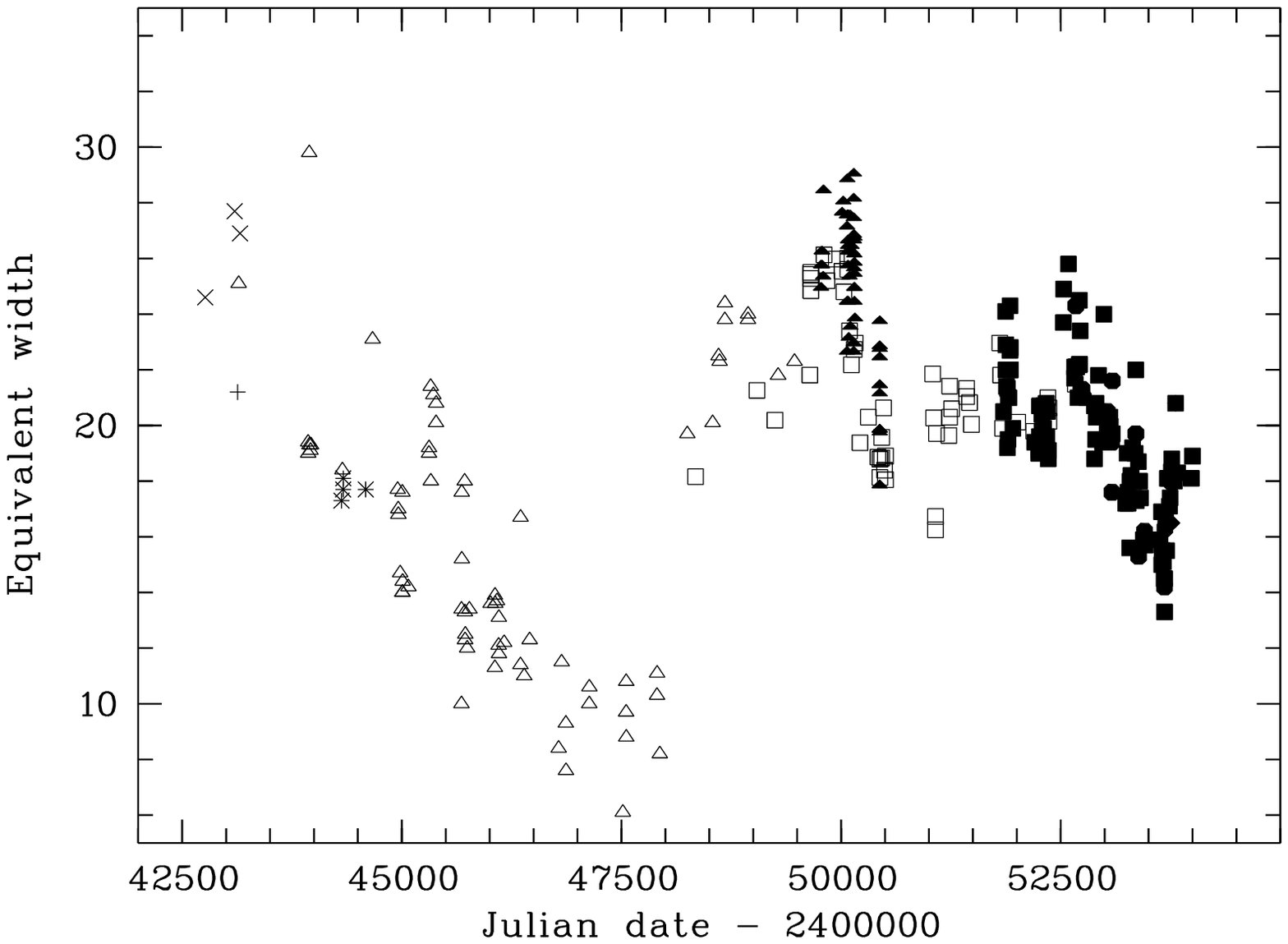}{H$\alpha$ equivalent widths of $\zeta$ Tau since
  1975. Data taken from the literature are plotted as open symbols: {\sc
    Heros} group (Rivinius et al., 2006, squares), Guo et al., 1995
  (triangles), Fontaine et al., 1982 (plus), Slettebak~\&~Reynolds, 1978
  (crosses), Andrillat~\&~Fehrenbach, 1982 (asterisks); data taken by various
  amateur observers as filled ones: Pollmann prism (filled triangle), Pollman
  grating (filled square), Stober (filled circles), and Schanne (filled
  diamonds).}

\IBVSfig{9cm}{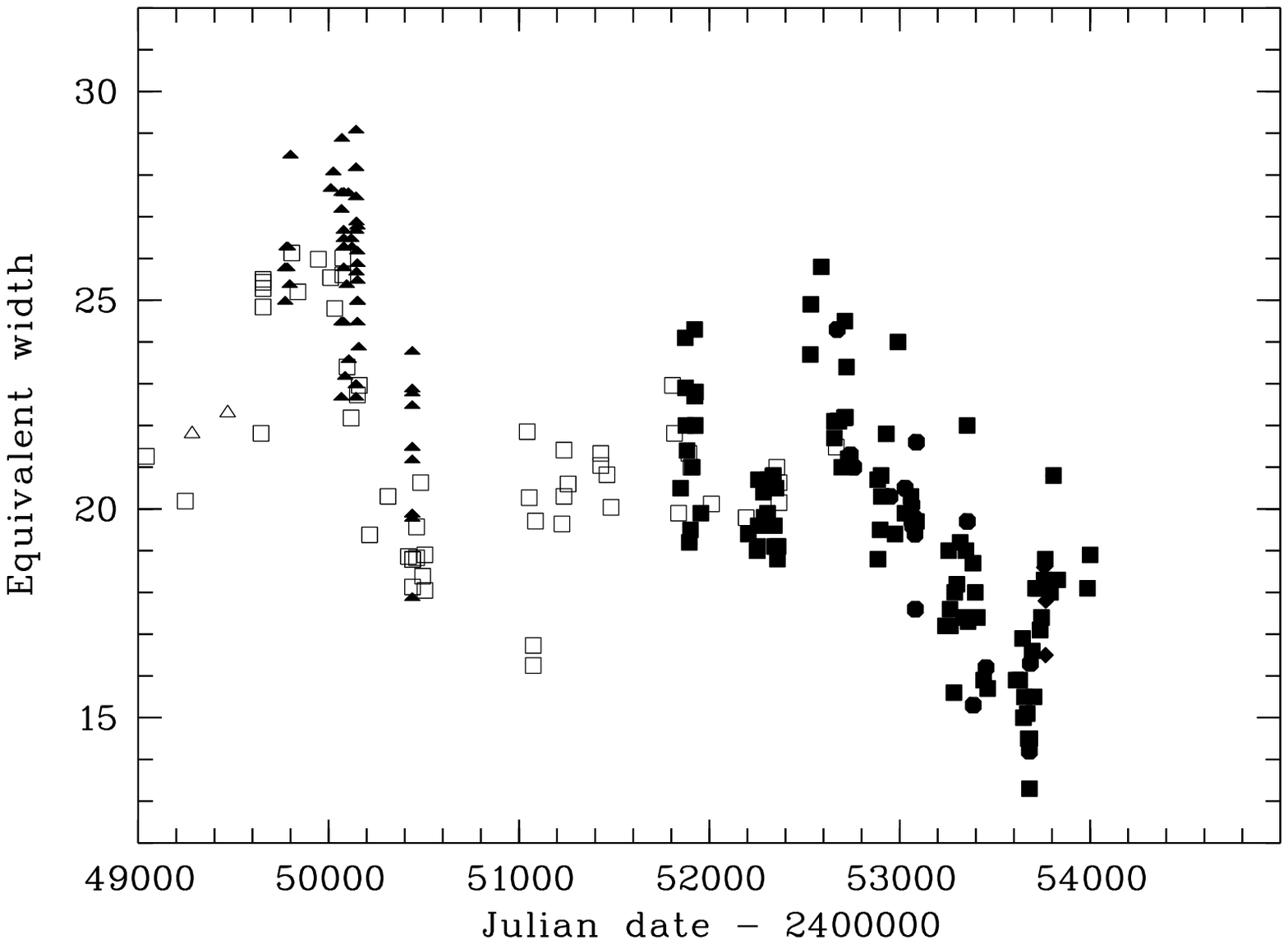}{Enlargement of Fig.\ 2 (see there for symbols and
  data sources), showing the data presented in this work in greater detail,
  also for comparison between values taken with professional and amateur
  equipment.}

Fig.~2 shows the measurements of this work combined with various published
values from about 1975 to 2006 to illustrate the longest variation time scale
present in $\zeta$ Tau, while Fig.\ 3 shows a closeup centered on the data
derived in this study. The EW currently is on a slow, but steady decline,
similar to the one seen before 1990.

\subsection{Peak height ratio}

The H$\alpha$-profile normally shows two emission peaks separated by a
central absorption core. In $\zeta$ Tau, both peaks strengths vary in
anti-phase respective to each other, so that the ratio of their violet
to red heights, called $V/R$-ratio, cyclically changes from $V>R$ to
$V<R$ and back. At times, however, the clear central absorption may
weaken or even disappear, and the emission peaks then may have
complicated appearance, split into sub-peaks and often called
triple-peak profile. The origin of such triple-peak profiles is
unclear. They generally appear at transitions from $V<R$ to $V>R$, but
not vice versa.  In the observations reported here such triple-peak
structures are seen from Dec.\ 2003 to Sept.\ 2004. The temporal
evolution of the H$\alpha$ profile between 2000 and 2006 is shown in
Fig.~1.

\IBVSfig{5.7cm}{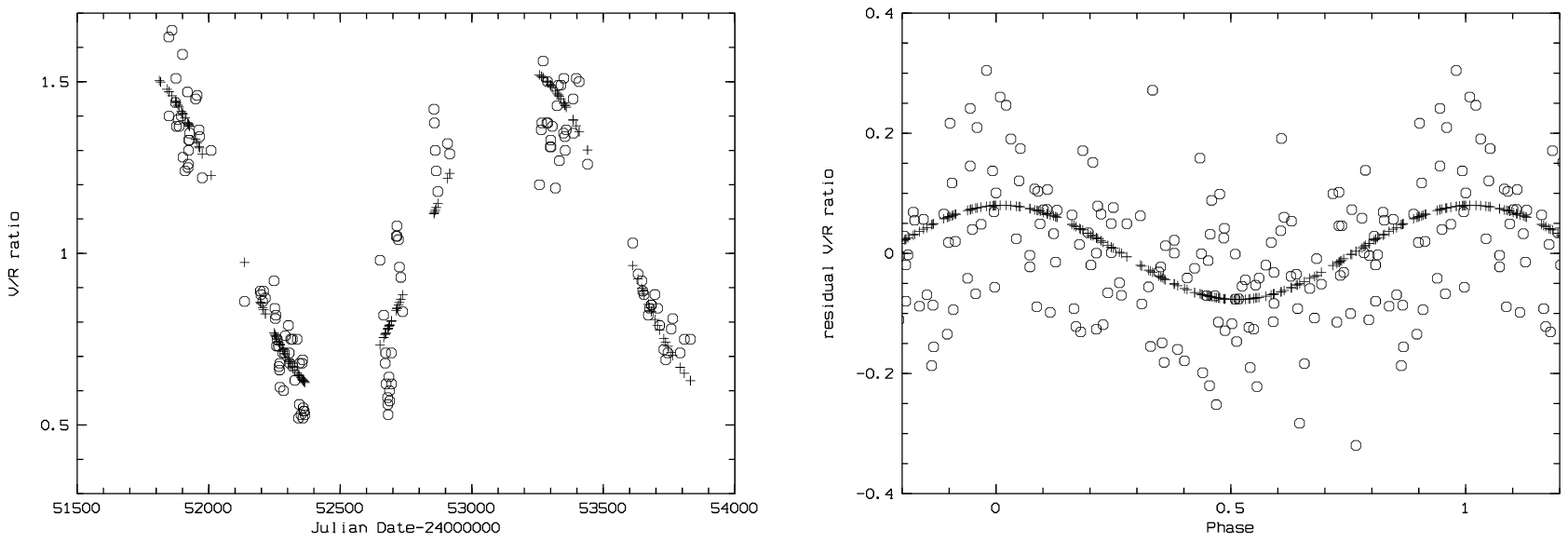}{H$\alpha$ $V/R$-ratio. Left: The measured
  values vs.\ Julian date (open symbols) and the sine wave with ${\cal
    P}=1471$\,d (plus signs). Right: The residuals of the left panel,
  folded with ${\cal P}=69.3$\,d and the respective sine fit. Shown
  are 1.4 cycles for clarification, i.e.\ 40\,\% of the points are
  redundant.}

$V/R$-ratio have been measured in the spectra in which both peaks are
apparent, and subjected to a formal period analysis using the time series
tools introduced by Kaufer et al. (1996). Note that the following
uncertainties are $1 \sigma$-errors. The first iteration reveals a $V/R$ cycle
time of 1471$\pm$15\,d, i.e.\ about 4.0\,years (Fig.~4, left).
While this is shorter than the 5 to 7 years in the list by Okazaki (1997)
derived from 1960 tom 1993, it is consistent with the 4.25 years cycle time
given by Rivinius et al.\ (2006) for 1991 to 2003. Given that only a
little more than one cycle is covered the main purpose of this exercise is to
pre-whiten the data for the analysis of shorter variations.

The second iteration on the residuals, i.e.\ after removing the sine wave fit
derived in the first step, reveals a 69.3$\pm$0.2\,d cycle (Fig.~4,
right). This cycle is clearly present during the  central part of the
dataset, but it is not of constant amplitude. The variance seen in the right
panel of Fig.~4 is well above the measuring uncertainty. In fact, looking at
individual seasons, the 69.3\,d cycle is not seen before JD=2\,452\,100,
hardly visible until 53000, but then becoming very strong, and finally
weakening again after JD=2\,453\,500. 

The ephemeris of the residual $V/R$ maximum is 
\[
2\,452\,996 + 69.3 \times E
\]

The cycle time of 69.3 days is about half of the orbital period of the
system of 132.97\,d (Harmanec, 1984), but a precise 1:2 ratio is well
outside a $3\sigma$ uncertainty.  As a check, sorting the data with
the orbital period rather shows the properties of a scatter diagram
than a meaningful phase curve.

Phase locking of the $V/R$ ratio has been observed in a number of
binaries.  However, while Harmanec et al.\ (2002) attribute this to
the property of the Roche lobe, e.g.\ for the case of 59\,Cyg,
\v{S}tefl et al. (2007, also Okazaki, priv. comm.) found in
hydrodynamical simulations that a true phase lock will not happen for
a density wave, usually thought to cause V/R variations. Rather, they
attribute precise locks, as in 59\,Cyg, to radiative effects (Maintz
et al., 2005) which is not likely in $\zeta$\,Tau, however.  Instead
of an exact tidal lock, the \v{S}tefl et al.\ mention that in
eccentric binaries tidally induced disturbances may develop with a
period slightly longer than the orbital one, and we may note that at
least the double-wave period would qualify under this statement.

This small difference may also offer an explanation for the strongly variable
amplitude: The orbital period, supposedly causing a tidal disturbance and the
$V/R$ variation cycle length as observed, would give rise to a long-term
beating period in the excitation mechanism of about 9 years.

\IBVSfig{16cm}{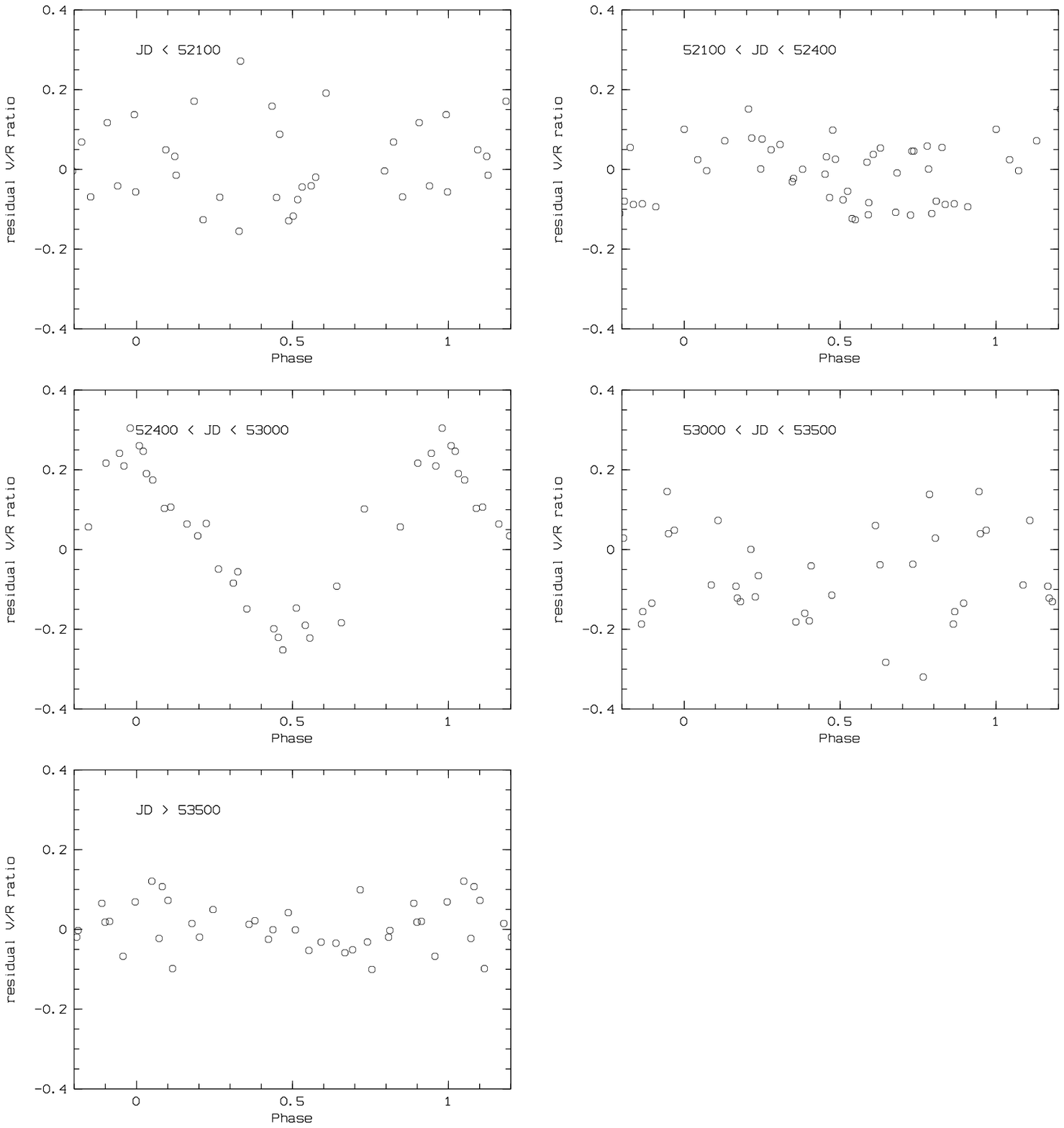}{Strength of 69.3\,d $V/R$-ratio cycles in
individual data subsets.}

\section{Discussion and Outlook}

The data presented in this work extend the $\zeta$ Tau spectra shown by
Rivinius et al., (2006, their Fig. A.4 in Appendix A). While their data cover
the years 1991 to 2003, the data here cover 2000 to 2006, with the observations
ongoing. 

Long-term spectroscopic monitoring by dedicated amateurs can deliver important
data for the professional community.  For instance, one easily recognizes
state of the $V/R$ cycle due to the one-armed density wave, as well as maxima
in equivalent width at 50150 and 52600, that do not coincide with the $V/R$
cycle.

The 69.3\,d cycle in spectroscopy is another example of a phenomenon almost
inaccessible to professional astronomers due to the observational timescales
required, which on the other hand poses no problem to the dedicated amateur
observer.

In the first few spectra of the 2006/2007 observing season, a sharp rise in
equivalent width from about 18 to 26\,\AA\ is seen. At the same time, as the
V/R ratio changes from  $V<R$ to $V>R$ again, the emission has developed a
triple-peak profile, entering a new cycle in its $V/R$ variations.
\bigskip
\bigskip
\bigskip

\noindent
Acknowledgenments:
\bigskip

\noindent
We are grateful to Petr Harmanec, the referee, who's detailed and
critical comments lead to major extensions and improvements of this
work.

\references 

Andrillat Y., Fehrenbach C., 1982, {\it A\&AS}, {\bf 48}, 93
\BIBCODE{1982A&AS...48...93A}

Fontaine G., Villeneuve B., Landstreet J.~D., Taylor R.~H., 1982, {\it ApJS},
{\bf 49}, 259 \BIBCODE{1982ApJS...49..259F}

Kaufer, A., Stahl, O., Wolf, B., et al., 1996,  {\it A\&A} {\bf 305}, 887 \BIBCODE{1996A&A...305..887K}

Guo, Y., Huang, L., Hao, J., Cao, H., Guo, Z., \& Guo, X.\ 1995, {\it A\&AS},
{\bf 112}, 201 \BIBCODE{1995A&AS..112..201G}

Harmanec, P.\ 1984, {\it Bulletin of the Astronomical Institutes of
 Czechoslovakia}, {\bf 35}, 164 \BIBCODE{1984BAICz..35..164H}

Harmanec, P., et al.\ 2002, {\it A\&A}, {\bf 387}, 580
\BIBCODE{2002A&A...387..580H}

Maintz, M., Rivinius, T., Stahl, O., Stefl, S., \& Appenzeller, I.\
2005, {\it Publications of the Astronomical Institute of the ASCR}
{\bf 93}, 21 \BIBCODE{2005PAICz..93...21M}

Okazaki, A.: 1997, {\it A\&A} {\bf 318}, 548 \BIBCODE{1997A&A...318..548O}

Rivinius, Th., \v{S}tefl, S, Baade, D.: 2006, {\it A\&A} {\bf 459}, 137
\BIBCODE{2006A&A...459..137R}

Slettebak A., Reynolds R.~C., 1978, {\it ApJS}, {\bf 38}, 205
\BIBCODE{1978ApJS...38..205S}

{\v S}tefl, S., Okazaki, A.~T., Rivinius, T., \& Baade, D.\ 2007, Active
OB-Stars: Laboratories for Stellar and Circumstellar Physics, {\it ASP Conf.\
Series} {\bf 361}, 274 \BIBCODE{2007ASPC..361..274S}

\endreferences
\end{document}